\documentclass[conference]{IEEEtran}
\IEEEoverridecommandlockouts
\usepackage{cite}
\usepackage{amsmath,amssymb,amsfonts}
\usepackage{graphicx}
\graphicspath{{./}}
\usepackage{textcomp}
\usepackage{xcolor}
\usepackage{booktabs}
\usepackage{multirow}
\usepackage{hyperref}
\usepackage{url}
\usepackage{newtxtext,newtxmath}
\usepackage{mdframed}
\usepackage{stfloats}
\mdfdefinestyle{leftbar}{
  topline=false,
  bottomline=false,
  rightline=false,
  innertopmargin=4pt,
  innerbottommargin=4pt,
  innerrightmargin=8pt,
  innerleftmargin=8pt,
  linewidth=2pt,
  linecolor=black!40
}
\newenvironment{leftbarquote}{\begin{mdframed}[style=leftbar]\small\itshape}{\end{mdframed}}
\def\BibTeX{{\rm B\kern-.05em{\sc i\kern-.025em b}\kern-.08em
    T\kern-.1667em\lower.7ex\hbox{E}\kern-.125emX}}
\makeatletter
\renewcommand{\subsubsection}{\@startsection{subsubsection}{3}{\z@}{2.0ex plus 1.0ex minus 0.2ex}{0.001ex}{\normalfont\normalsize\itshape}}
\makeatother
\begin{document}

\title{The Virtual Roundtable: Multi-Agent Personas Simulating the Dynamics of Human Brainstorming}

\author{
\IEEEauthorblockN{Tim Dorn\IEEEauthorrefmark{1}*, Saara A. Khan\IEEEauthorrefmark{1}, Julie Mumford\IEEEauthorrefmark{1}}
\IEEEauthorblockA{\IEEEauthorrefmark{1}\textit{Reality Labs}, \textit{Meta}, Menlo Park, California 94025, USA \\
*Corresponding Author}
}

\maketitle

\begin{abstract}
As AI-driven product development accelerates, the bottleneck is shifting from how we build to what we build. Traditional human brainstorming faces challenges including groupthink, echo chambers, and limited diversity. To address this, we present a multi-agentic architecture that simulates roundtable brainstorming through two phases: divergent thinking to generate diverse ideas, and convergent thinking to evaluate and rank the most promising ones. The system employs diverse AI personas that engage in roundtable discussions, guided by an agentic facilitator that steers the discussion toward productive outcomes. Personas maintain private thoughts while commenting publicly, with ideas emerging organically throughout the discussion. Per-persona quotas on idea submissions and votes promote balanced participation while producing natural rankings. Throughout the session, the system tracks each idea's lineage, capturing how concepts originate and cross-pollinate over time. We demonstrate this approach through a case study generating consumer ideas for AI smart glasses, showing (i) it produces diverse, relevant ideas with insights into their evolution; (ii) the cumulative exchange of perspectives across personas cultivates a shared context that progressively deepens the quality of discussion and the ideas produced.
\end{abstract}

\section{Introduction}
The proliferation of Large Language Models (LLMs) has dramatically accelerated the software development lifecycle. As the ability to rapidly build and iterate becomes commoditized, a new bottleneck emerges: knowing what to build. The cognitive burden shifts from implementation to ideation, requiring methodologies that can quickly generate and evaluate high-quality, customer-aligned product concepts~\cite{b1}.

Traditional brainstorming and customer focus groups, cornerstones of human-centered design~\cite{b2}, offer a proven model for this process, yet suffer from inherent limitations: susceptibility to conformity, logistical complexity, and idea diversity constrained by participant recruitment~\cite{b3,b4}. To address these challenges, we developed a multi-agent AI architecture modeled on the dynamics of a roundtable brainstorming session. The system operationalizes Osborn's four foundational rules of effective brainstorming~\cite{b7}---no bad ideas, strive for quantity, build on others' ideas, and welcome unconventional thinking---while structurally separating divergent idea generation from convergent idea evaluation~\cite{bGuilford}. An agentic facilitator monitors session dynamics and intervenes to guard against groupthink, balance participation, and steer the group toward productive outcomes. This mirrors the natural rhythm of how effective groups collectively move from exploring a problem to ranking solutions.

Concretely, our architecture orchestrates a facilitated conversation among a diverse group of AI personas, each defined by a unique character sketch, around a given brainstorming topic. Personas operate across three cognitive layers: private thoughts for internal reflection, public comments for group dialogue, and public ideas for formal proposals, mirroring how humans balance inner reasoning with outward communication. Every contribution is linked to the artifacts that inspired it, forming a weighted influence graph that traces how concepts flow, merge, and evolve.

Our objectives for this paper are twofold: (i) demonstrate the feasibility of such a multi-agent architecture for brainstorming, and (ii) investigate how sustained cross-persona discussion affects the depth of cross-pollination and the diversity of resulting ideas.

\begin{figure*}[t]
\centerline{\includegraphics[width=0.95\textwidth]{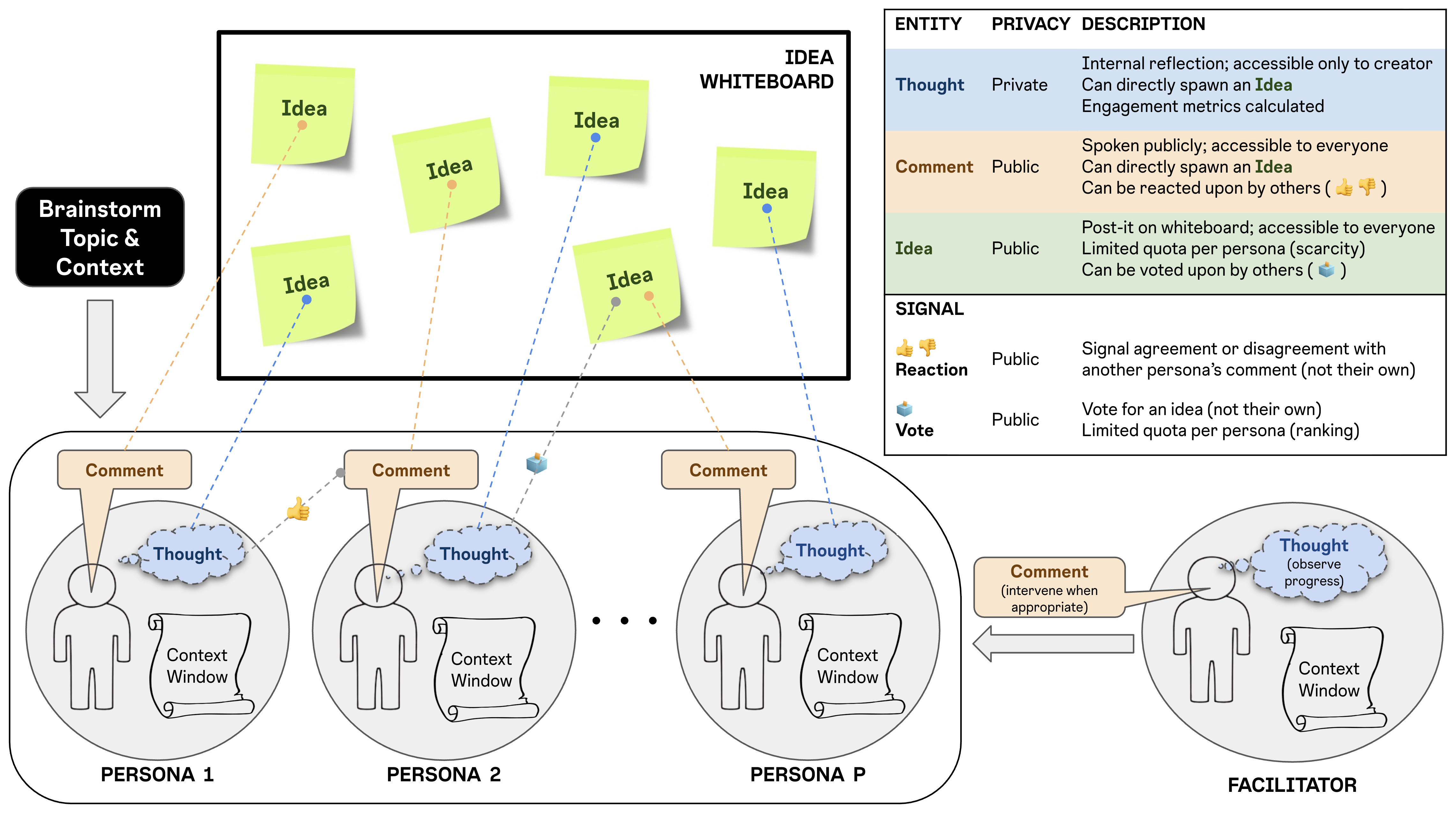}}
\caption{Four foundational abstractions of the brainstorming architecture: Personas participate in the session, Entities capture discourse artifacts (thoughts, comments, ideas), Signals express agreement through reactions and votes, and a Facilitator steers session progression.}
\label{fig:abstractions}
\end{figure*}

\section{Related Work}
Our work builds on established human-centered design principles and extends them into a multi-agent AI context. Foundational to our approach is the practice of brainstorming, which emphasizes divergent idea generation followed by convergent evaluation~\cite{b7}. Research confirms that heterogeneous groups produce more diverse and creative solutions by cross-pollinating varied perspectives~\cite{b8,b9}. This is complemented by customer focus groups, a qualitative method used to discover what users want from a system and why~\cite{b10}. Focus groups are highly effective for both generative research, by uncovering novel ideas through problem detection~\cite{b11}, and for evaluative research, by validating early concepts to ensure they align with customer expectations.

Recent work has shown that (i) multi-agent LLM systems can improve factuality and reasoning through iterative refinement among agents~\cite{b5}; and (ii) role-play prompting can enhance individual agent responses~\cite{b14}. CAMEL~\cite{b13} introduced cooperative role-playing between two domain-specialized agents for task completion, and Wang et al.~\cite{bWang} extended this to multi-persona self-collaboration. Our architecture extends this line of research to open-ended ideation by orchestrating a diverse group of persona agents in a facilitated brainstorming session that simulates how humans collaboratively generate and evaluate ideas.

\section{System Architecture}

\subsection{Foundational Abstractions}
Our architecture models the dynamics of brainstorming through four core abstractions (Fig.~\ref{fig:abstractions}). \textbf{Personas} represent individual agents participating in the session. \textbf{Entities} capture discourse artifacts, including private thoughts, public comments, and public ideas. \textbf{Signals}---comprising reactions and votes---enable participants to express agreement with comments or formally endorse ideas. Finally, a \textbf{Facilitator} is a specialized agent that contributes no ideas of its own but instead monitors session dynamics, guarding against groupthink, managing dominant voices, encouraging quieter participants, and steering the group from divergent idea generation to convergent evaluation~\cite{b3,b16}.

A central \textbf{Event Bus} mediates all communication among these components. Rather than agents interacting directly, every discourse action is published as an event onto a shared channel, with the bus controlling which events each agent receives. This enforces a layered visibility model: (i) comments, ideas, reactions, and votes are public events broadcast to all personas; (ii) thoughts remain private to their originating agent; and (iii) facilitator interventions are routed only to their intended recipients. This separation ensures each persona receives only the context relevant to its deliberation.

\subsection{Persona Generation}
To ensure consistency in persona generation, each persona's character sketch was generated using the following prompt:

\begin{leftbarquote}
``Craft a sharp, vivid character sketch---not a list of traits. In a few sentences, bring this persona to life with a distinct perspective, voice, and way of engaging with the world. Describe who they are, what they care about, and how they think. Write approximately 60--80 words. Every sentence should sharpen what makes this persona distinct, unique, and specialize them in their particular field.''
\end{leftbarquote}

Each sketch captures the persona's expertise, values, and worldview, and is embedded in all subsequent prompts to maintain a consistent perspective throughout the session. While outside the scope of this paper, we note that persona generation is itself an active area of research, as the quality of the persona's sketch is likely to influence the quality of ideas produced during brainstorming~\cite{b6,b17,b18,bWang}.

\subsection{Brainstorming Phases: Discussion, Ideation, Voting}
\begin{figure}[t]
\centering
\includegraphics[width=\columnwidth]{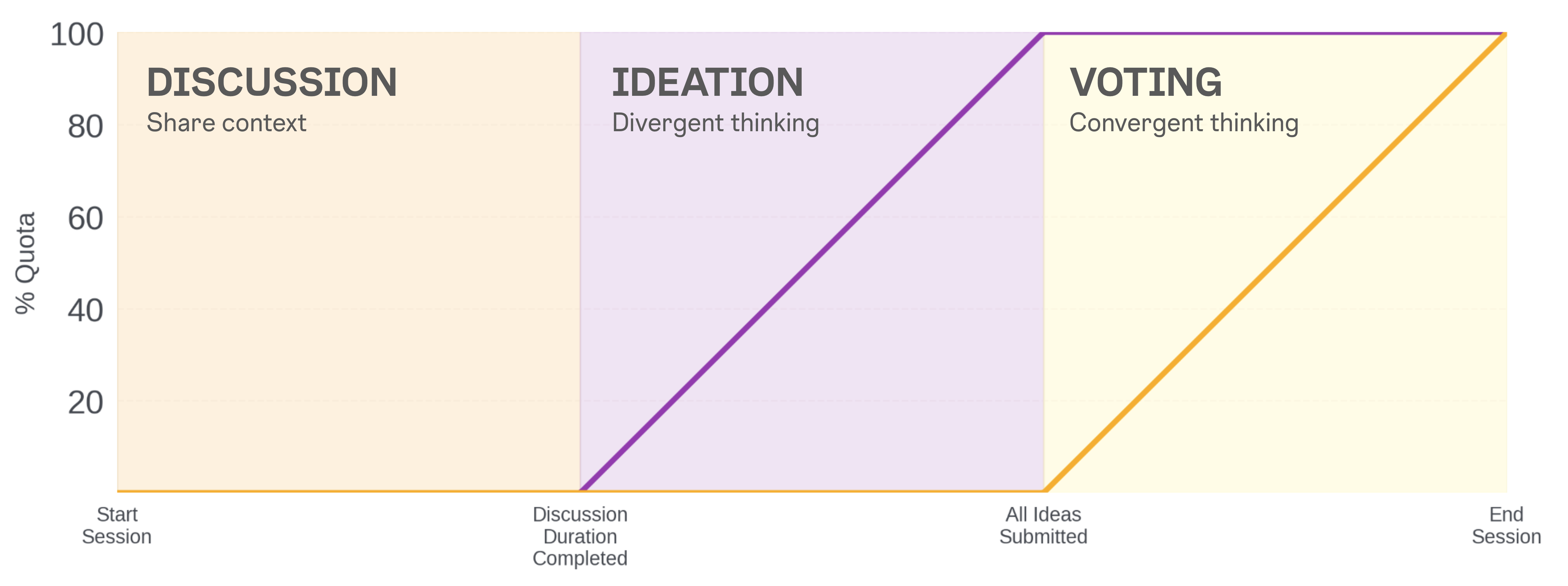}
\caption{Expected trajectory of idea submissions (purple) and votes (amber) as a percentage of quota across the three phases. Vertical bands indicate phase boundaries.}
\label{fig:pacingphasesexpected}
\end{figure}

The brainstorming session proceeds through three phases---\textbf{Discussion}, \textbf{Ideation}, and \textbf{Voting}---each serving a complementary cognitive function (see Fig.~\ref{fig:pacingphasesexpected}).

\begin{table*}[b!]
\centering
\caption{Phase-dependent context provided to each persona's system prompt.}
\label{tab:phasecontext}
\footnotesize
\begin{tabular}{p{0.12\textwidth} p{0.26\textwidth} p{0.26\textwidth} p{0.26\textwidth}}
\toprule
\textbf{Context} & \textbf{Discussion Phase} & \textbf{Ideation Phase} & \textbf{Voting Phase} \\
\midrule
\textbf{Persona \& Topic} & \multicolumn{3}{p{0.78\textwidth}}{Persona identity and description, brainstorm topic, list of other personas participating in the brainstorm session} \\
\midrule
\textbf{Phase Guidance} & \textit{``Discuss the topic openly''} & \textit{``Submit ideas if you're inspired''} & \textit{``Cast votes if one resonates with you''} \\
\midrule
\textbf{Ideas Published} & Empty & Ideas ordered by recency & Ideas ordered by vote count then recency. Own ideas are marked as non-votable \\
\midrule
\textbf{Discussion History} & \multicolumn{3}{p{0.78\textwidth}}{All comments + own private thoughts, interleaved chronologically (most recent first). Own entities are explicitly marked. Entities with high reaction counts are explicitly marked.} \\
\midrule
\textbf{Participation Metrics} & \multicolumn{3}{p{0.78\textwidth}}{Per-persona breakdown of comment count, comment share \%, idea quota exhausted, vote quota exhausted. If there are any targeted facilitator interventions for the personas, they are surfaced here.} \\
\midrule
\textbf{Entity Outputs} & Thought, Comment & Thought, Comment, Idea* & Thought, Comment \\
\textbf{Signal Outputs} & Thumbs Up*, Thumbs Down* & Thumbs Up*, Thumbs Down* & Thumbs Up*, Thumbs Down*, Vote* \\
\textbf{Lineage Output} & \multicolumn{3}{p{0.78\textwidth}}{Explicit references to the prior discussion entities that influenced the current Thought or Comment; each entity is uniquely identifiable in the Discussion History} \\
\bottomrule
\end{tabular}
\par\vspace{6pt}
\raggedright\scriptsize *Optional outputs, emitted agentically by the persona during a Thought or Comment.
\end{table*}

The session opens with a \textbf{Discussion} phase, in which personas engage in open dialogue to frame the problem, surface assumptions, and build shared context. No ideas are captured and no evaluative judgments are solicited---what Sawyer~\cite{bSawyer} terms ``problem finding.'' Both idea and vote trajectories remain at zero throughout.

Once shared understanding is established, the \textbf{Ideation} phase shifts the group toward divergent idea generation. Personas generate ideas agentically until the target quota is met, while voting remains suppressed. This separation of generation from evaluation operationalizes Osborn's (1953) principle of deferred judgment, broadening creative output while mitigating anchoring, production blocking, and premature convergence~\cite{b3, b4}.

Only after all ideas have been submitted does the \textbf{Voting} phase begin. Each persona evaluates the complete idea set according to its values and worldview, casting votes agentically until its quota is met. These per-persona quotas for ideas and votes serve a dual purpose: they prevent any single voice from dominating the session, and they transform the voting phase into a natural ranking mechanism.

Because quotas make the ideation and voting phases self-regulating, the discussion phase duration becomes the primary design lever for session length. We hypothesize that extending discussion will yield higher-quality ideation by giving personas more time to cross-pollinate context before idea generation begins. This draws on the finding that exposure to others' viewpoints increases both the diversity and originality of subsequently generated ideas~\cite{bPaulusBrown}---a longer discussion surfaces a wider range of frames, analogies, and domain knowledge, giving participants richer material to recombine into novel ideas~\cite{bMednick}.

\subsection{The Session Loop}

\begin{figure*}[t]
\centerline{\includegraphics[width=\textwidth]{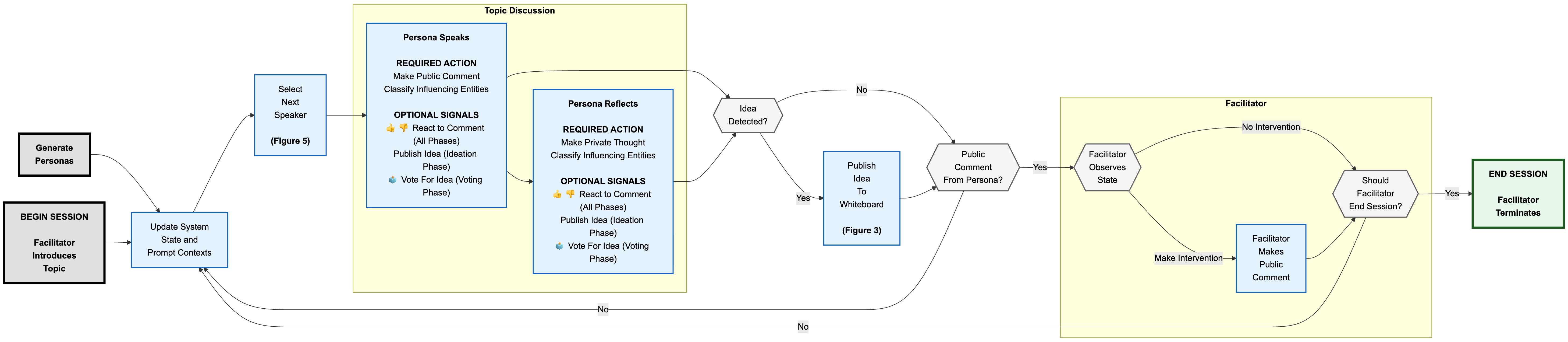}}
\caption{Event-driven execution loop of the multi-agent brainstorming system. After persona generation, the discussion proceeds cyclically through public comments, private thoughts, and speaker selection, with ideas and votes emerging throughout.}
\label{fig:execution}
\end{figure*}

During all phases, the session operates through an event-driven loop (Fig.~\ref{fig:execution}). The facilitator opens by introducing the topic and selecting an initial speaker at random. The discussion then proceeds cyclically: (i) a persona delivers a public comment, (ii) the remaining personas generate private thoughts in response, and (iii) the next speaker is selected.

The context and actions available to each persona vary by phase, as detailed in Table~\ref{tab:phasecontext}. Across all phases, personas may react to prior comments with thumbs up/down signals. During the ideation phase, personas may publish new ideas to a shared whiteboard, and once the voting phase begins, personas may cast votes for existing ideas. The facilitator ends the session once all idea and vote quotas have been fulfilled. Several elements of this loop warrant closer examination, and we discuss them below.

\subsubsection{Dialogue Flow \& Influence Tracking} As in human brainstorming, personas contribute through two complementary channels: private thoughts and public comments. Comments proceed sequentially, and each one triggers concurrent private reflections across all other personas, interpreted through their own respective character sketches. Ideas, votes, and thumbs up/down reactions are not independently elicited but arise organically within thoughts and comments as part of the generative process. Finally, each thought and comment explicitly cites the prior entities that informed it, enabling a full reconstruction of an idea's lineage.

\subsubsection{Idea Publication} 
\begin{figure}[t]
\centering
\includegraphics[width=\columnwidth]{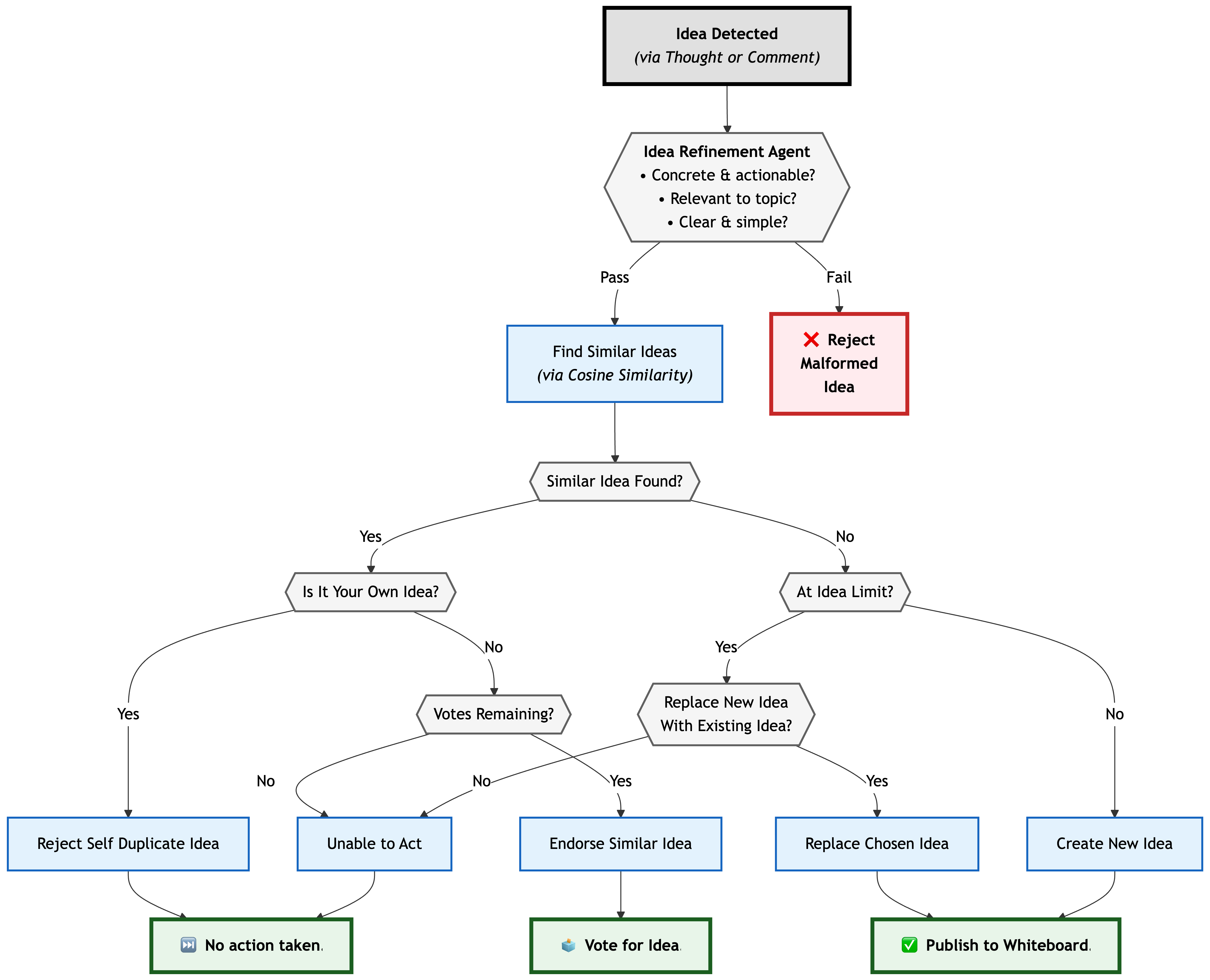}
\caption{Idea Submission. When a potential idea surfaces from a persona, it undergoes a refinement process that evaluates topic adherence and similarity to existing ideas before being published to the whiteboard.}
\label{fig:ideaflow}
\end{figure}

When an idea surfaces from a persona's thought or comment, it undergoes a lightweight refinement process, not to evaluate the idea itself, but to check for linguistic clarity, topical relevance, and duplication. If a similar idea already exists on the board, as determined by the cosine similarity of their relative embeddings~\cite{bDrama} exceeding a threshold, a vote is cast for the existing idea rather than creating a duplicate. While votes are ordinarily restricted to the Voting Phase, deduplication-triggered votes are the sole exception, capturing the signal of idea similarity while keeping the whiteboard free of redundancy. This process promotes a broad and diverse collection of unique ideas on the whiteboard while mitigating the risk of off-topic idea hallucination (Fig.~\ref{fig:ideaflow}).

\begin{figure}[t]
\centering
\includegraphics[width=\columnwidth]{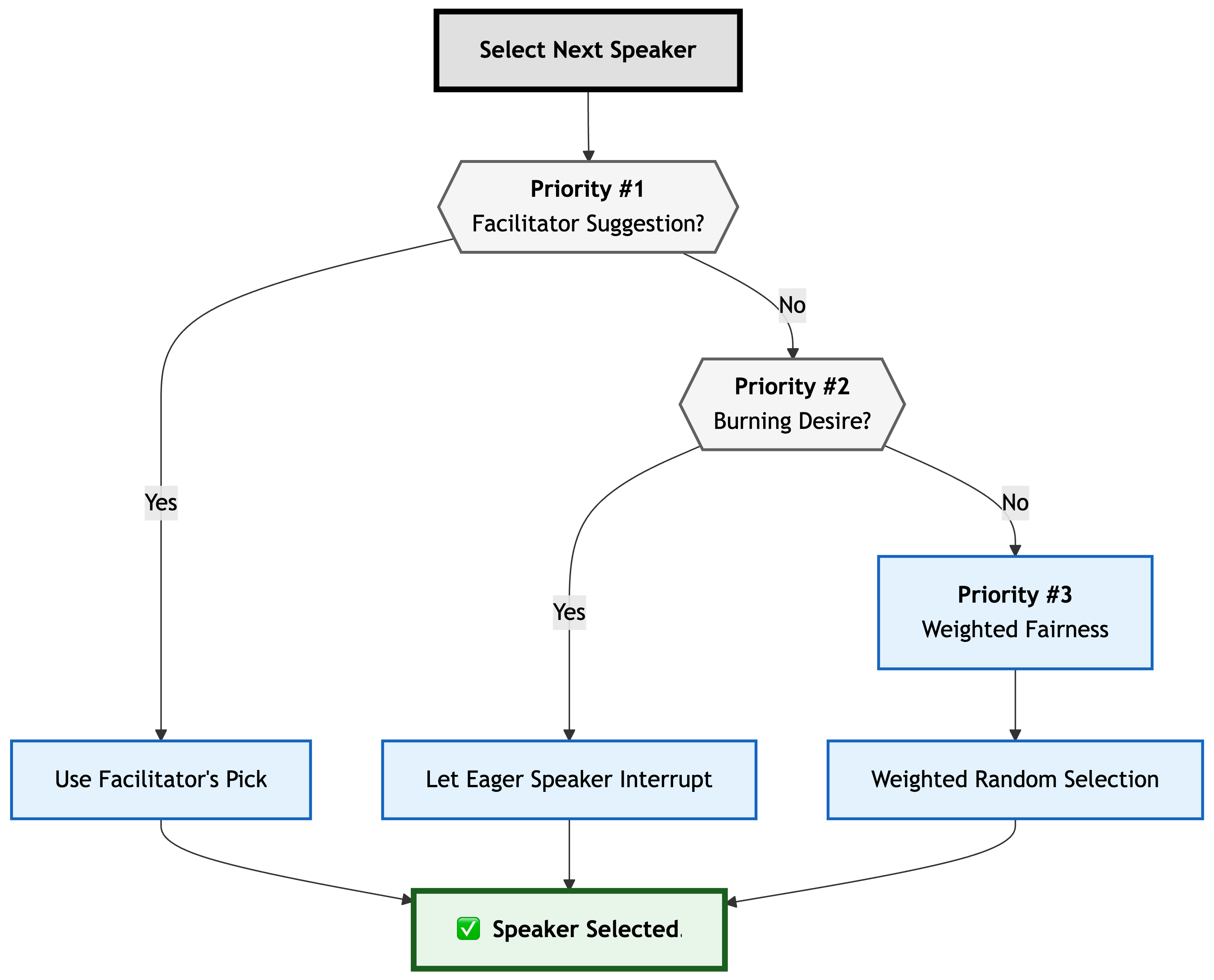}
\caption{Speaker Selection. A three-tier priority system determines the next speaker: facilitator designation, eager-speaker interruption, or weighted fairness based on persona eagerness-to-speak and participation volume.}
\label{fig:speakerflow}
\end{figure}

\subsubsection{The Facilitator} The facilitator is a specialized agent that operates outside the topical discussion, contributing no ideas of its own. After each persona comment it observes a session snapshot of aggregated and per-persona metrics similar to that shown in Table~\ref{tab:phasecontext}, and evaluates whether the conversation should proceed unguided or requires an explicit intervention. Its mandate is to guard against process loss---groupthink, unbalanced participation, echo chambers, and anchoring~\cite{b16,b3}---under a restraint-first philosophy in which silent observation is the default. When the facilitator chooses to intervene, it selects both the target persona(s) and the specific guidance to issue to course correct.

This creates a closed feedback loop between the two types of agents (Fig.~\ref{fig:execution}). Facilitator interventions are embedded as additional context in the targeted personas' future prompt (Table~\ref{tab:phasecontext}), each tagged with an age marker recording how many comments have elapsed since issuance. Personas interpret these instructions through the lens of their character sketches and agentically decide when to act on them. When they do, they mark the intervention as ``responded,'' closing the loop. Unresolved interventions persist in the facilitator's and targeted personas' contexts until resolved, preventing duplicate interventions and ensuring compliance.

\subsubsection{Speaker Selection} Speaker selection follows a three-tier priority system that mimics natural conversational dynamics (Fig.~\ref{fig:speakerflow}). At the highest priority level, the facilitator may explicitly designate the next speaker. Alternatively, a persona with a high ``eagerness-to-speak'' score, calculated during every private thought, may interrupt to claim the floor---akin to someone fervently raising their hand wanting to speak. In the absence of either condition, a weighted selection algorithm determines the next speaker based on eagerness, volume of participation, and a stochastic element, ensuring balanced contribution across all personas.

\begin{figure*}[t]
\centerline{\includegraphics[width=0.95\textwidth]{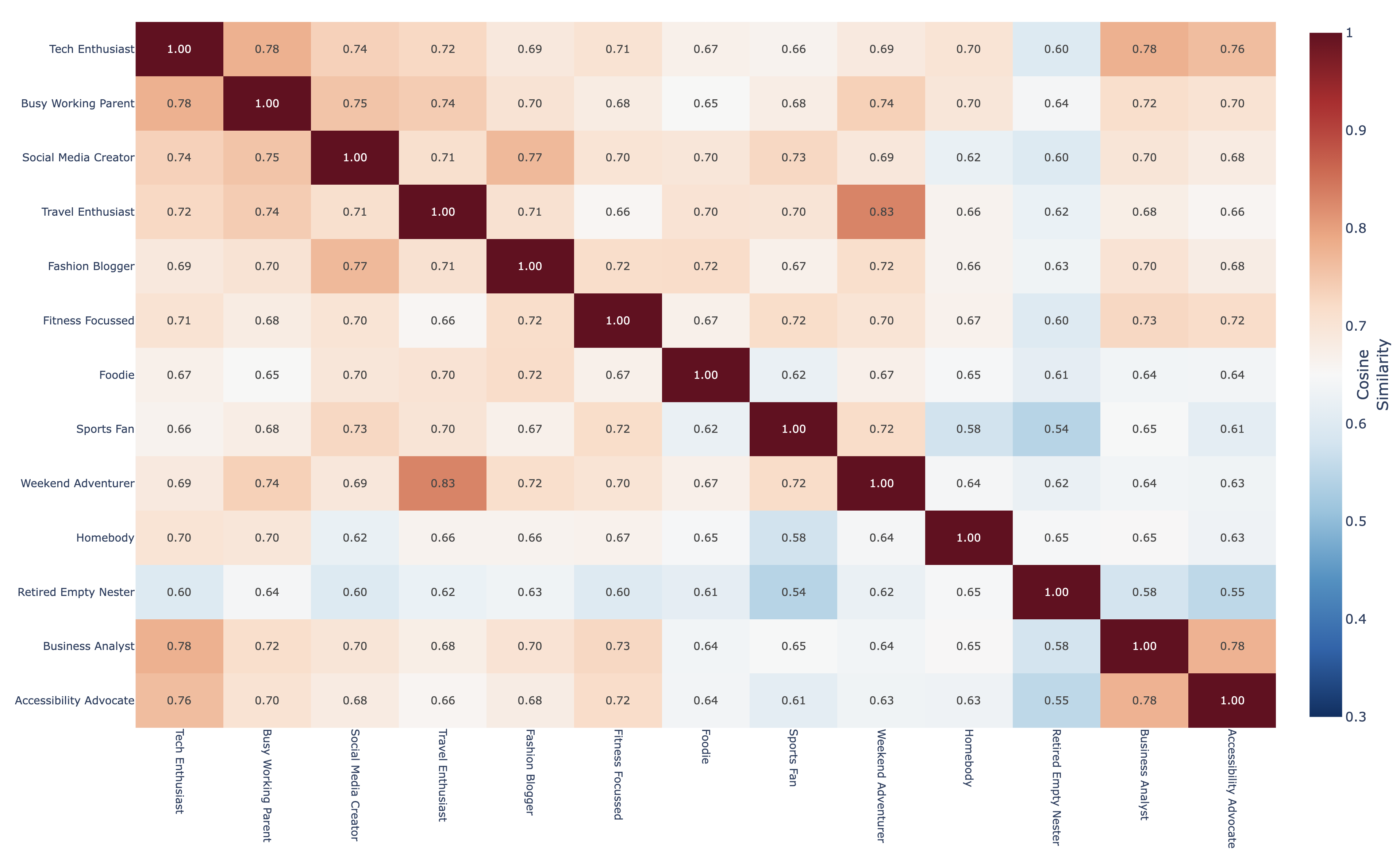}}
\caption{Pairwise cosine similarity matrix of the 13 AI persona character sketches. Lower scores (blue) indicate more distinct personas, while higher scores (red) indicate more similar persona sketches.}
\label{fig:personas}
\end{figure*}

\begin{figure*}[t]
\centerline{\includegraphics[width=0.95\textwidth]{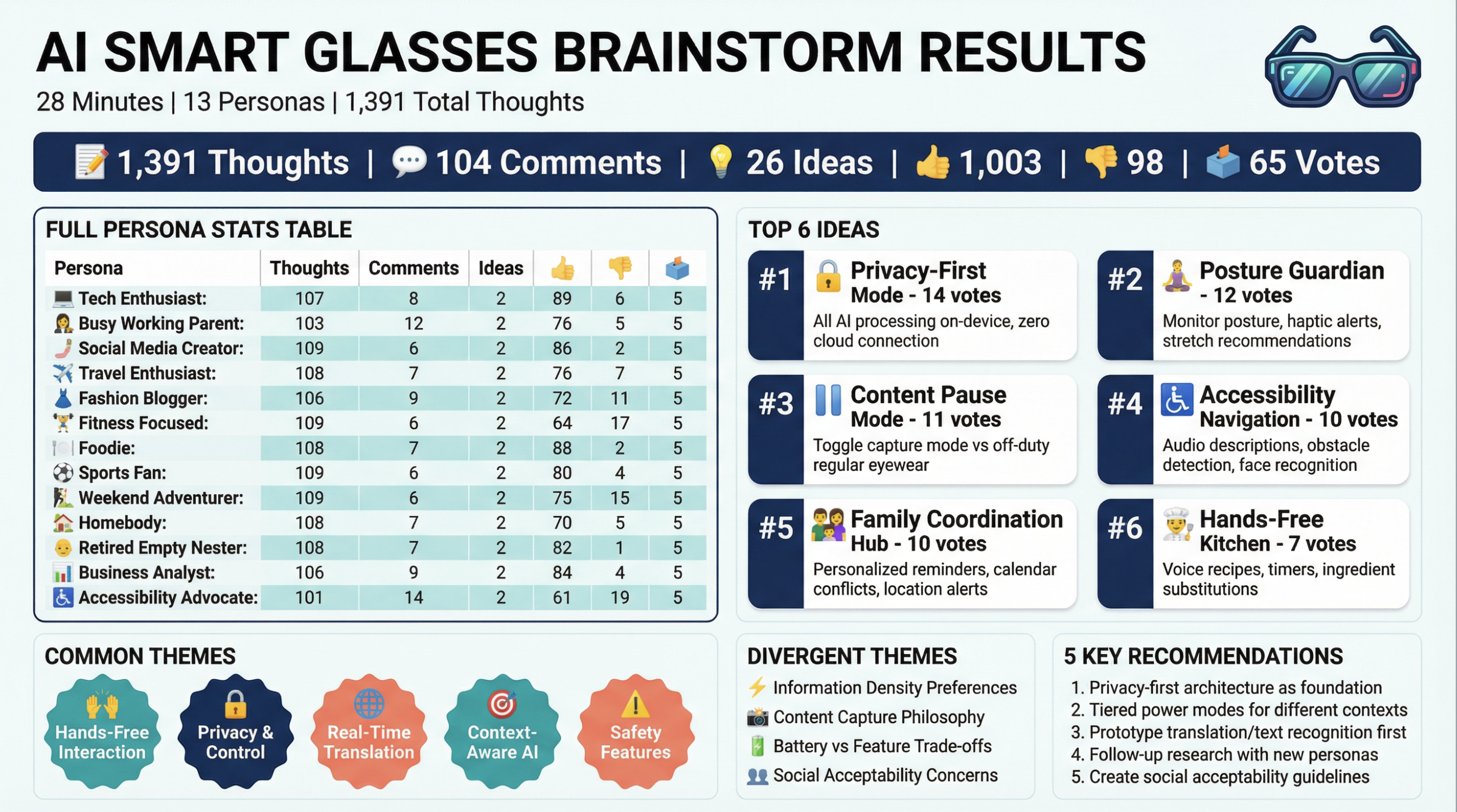}}
\caption{Synthesized brainstorming session output for the AI smart glasses case study, showing ranked ideas with vote counts, per-persona contributions, and the developmental lineage of each idea.}
\label{fig:output}
\end{figure*}

\section{Objective 1: Feasibility of a Facilitated Multi-Agent Architecture for Brainstorming}

\subsection{Persona Outputs}
We generated 13 AI personas---Tech Enthusiast, Busy Working Parent, Social Media Creator, Travel Enthusiast, Fashion Blogger, Fitness Focused, Foodie, Sports Fan, Weekend Adventurer, Homebody, Retired Empty Nester, Business Analyst, and Accessibility Advocate---using the following base prompt:

\begin{leftbarquote}
``Craft a sharp, vivid character sketch---not a list of traits. In a few sentences, bring this persona to life with a distinct perspective, voice, and way of engaging with the world. Describe who they are, what they care about, and how they think. Write approximately 60--80 words. Every sentence should sharpen what makes this persona distinct, unique, and specialize them in their particular field.''
\end{leftbarquote}

As an illustrative example, the character sketch produced for the Busy Working Parent persona was:

\begin{leftbarquote}
``You are a Busy Working Parent. You are a master of ruthless prioritization who mentally triages every moment between client deadlines and soccer practice pickups, viewing efficiency as both survival skill and art form. You think in overlapping timelines—mentally drafting emails while packing lunches, negotiating conference calls around school holidays, and measuring success not just in quarterly results but in being present for the moments that matter. Your phone buzzes with Slack notifications and permission slip reminders with equal urgency, and you've perfected the delicate choreography of professional competence and parental devotion.''
\end{leftbarquote}

As illustrated in Fig.~\ref{fig:personas}, pairwise cosine similarity scores among the 13 persona character sketches~\cite{bDrama} ranged from 0.54 to 0.83 ($M = 0.68$, $SD = 0.05$), indicating that meaningfully differentiated personas were generated to participate in the brainstorm session. The highest-similarity pair, Travel Enthusiast and Weekend Adventurer (0.83), shared an obvious thematic overlap in exploration and novelty-seeking, while at the other extreme, Sports Fan and Retired Empty Nester (0.54), exhibited the weakest overlap, reflecting divergent lifestyles and priorities. More broadly, professionally oriented personas (Tech Enthusiast, Business Analyst, Accessibility Advocate) formed a loose cluster with inter-similarities of 0.70--0.78, while domestically anchored personas (Homebody, Retired Empty Nester) sat apart, with consistently lower similarity to the rest of the personas. This structured diversity is not incidental: it ensured that the subsequent brainstorming session drew on a wide spectrum of perspectives, yielding the heterogeneous outputs we explore in the following section.

\subsection{Brainstorm Output}
\begin{figure}[t]
\centering
\includegraphics[width=\columnwidth]{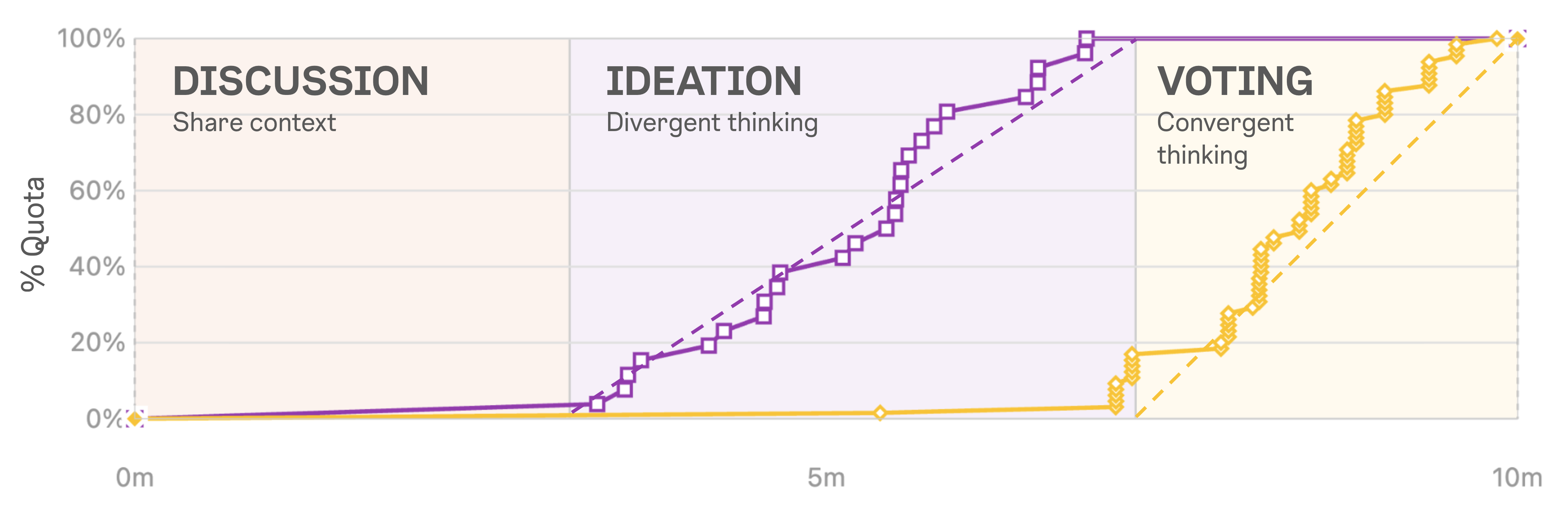}
\caption{Actual (solid line) compared to expected (dashed line) trajectory of paced idea submissions (purple) and votes (amber) as a percentage of quota across the three phases. Vertical bands indicate phase boundaries.}
\label{fig:pacingphasesactual}
\end{figure}

The 13 AI-generated personas were presented with a central brainstorming prompt: ``How might we utilize AI smart glasses to make our life easier and more productive?'' Each persona could submit up to 2 ideas and was allocated 5 voting credits. The discussion phase duration was set to 3 minutes, after which ideation and voting phases proceeded agentically until all quotas were fulfilled. In total, the session ran for approximately 10 minutes (Fig.~\ref{fig:pacingphasesactual}), producing a chronological transcript of thoughts, comments, and ideas that built iteratively on one another under the facilitator's guidance. Fig.~\ref{fig:output} distills the session into a single infographic.

\subsection{Illustrating Brainstorming Dynamics With Examples}
A central question we wanted to investigate is whether the resulting discussion produces genuine deliberation or merely a collection of independent suggestions. In this section, we examine the dynamic interplay that emerged during a brainstorming session.

\subsubsection{Idea Evolution Through Debate}
The top-voted idea of the session, ``Privacy-First Mode,'' illustrates how debate shaped proposals (Fig.~\ref{fig:output}). The concept did not emerge fully formed. The Tech Enthusiast initially proposed a hybrid cloud/edge processing architecture for smart glasses, optimized for battery life, a technically sound but narrow solution. It took sustained pushback from multiple personas to force a fundamental rethink. The Tech Enthusiast's own lineage summary of their idea captures this shift:

\begin{leftbarquote}
{\upshape [Tech Enthusiast]:} ``My second idea emerged from a crucial tension that developed throughout the brainstorm: I initially proposed hybrid cloud/edge processing to achieve longer battery life, but Homebody, Retired Empty Nester, and Travel Enthusiast powerfully challenged this approach, arguing that privacy and offline reliability aren't negotiable trade-offs. Their concerns about surveillance, data control, and functionality in remote areas forced me to completely reconsider my technical assumptions. This led to the `Privacy-First Local Mode' concept---fully on-device processing with zero cloud dependency.''
\end{leftbarquote}

What began as a technical optimization for longer battery life was confronted with user-centric values---privacy, reliability, autonomy---and the outcome was not a watered-down compromise but a fundamentally stronger proposal born from the collision of competing priorities.

\subsubsection{Persona Influence}
This pattern of influence extended beyond individual idea refinement. In several cases, one persona's core values directly reframed how another persona thought about the problem. For example, the Accessibility Advocate was a particularly powerful catalyst, consistently anchoring the conversation in principles of universal design. The Travel Enthusiast's lineage summary for their ``Accessibility Navigation'' idea (Fig.~\ref{fig:output}) shows this reframing in action:

\begin{leftbarquote}
{\upshape [Travel Enthusiast]:} ``The Accessibility Advocate's powerful intervention reframed everything, making me understand that the features I wanted for travel (navigation, translation, scene recognition) were actually foundational accessibility needs, not lifestyle upgrades. This led to my second, more refined idea\ldots{} combining the practical travel needs I'd identified with the universal design principles and `trustworthy first, optimized second' philosophy that emerged from the group's collective wisdom.''
\end{leftbarquote}

This demonstrates how a specialized viewpoint can enrich the broader discussion, turning a feature from niche to essential, ultimately reinforcing the product's foundation.

\begin{figure*}[!b]
\centerline{\includegraphics[width=\textwidth]{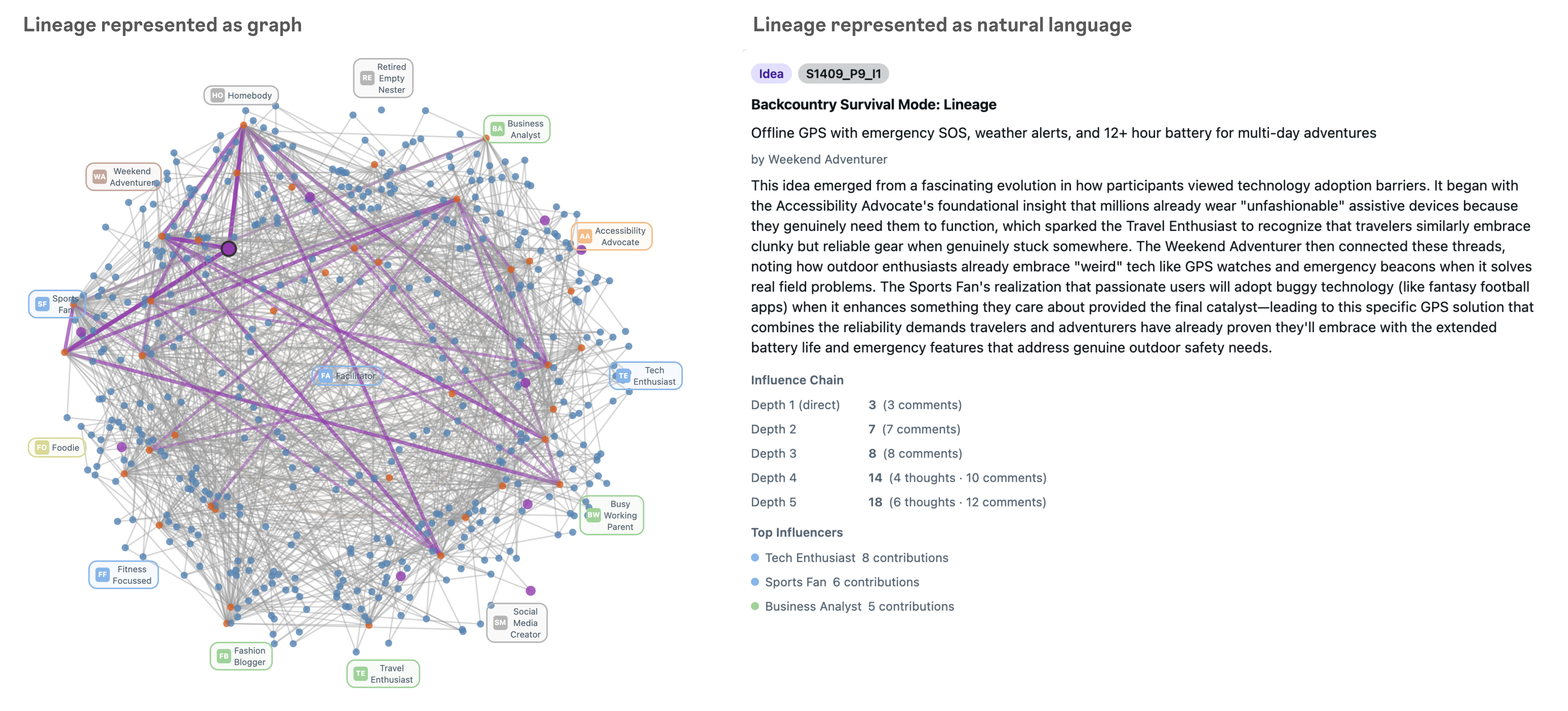}}
\caption{Lineage of a representative idea, shown as a graph (left) and as structured natural language (right). In the graph, persona nodes are linked by citation-based edges to influencing thoughts (blue), comments (orange), and ideas (purple); the highlighted path from the target idea (large purple node) traces its upstream influences, with edge thickness encoding proximity, thicker lines for closer contributions and thinner lines for more distant ones.}
\label{fig:influence}
\end{figure*}

\begin{table*}[b]
\centering
\caption{Metrics used to evaluate the effect of discussion duration on brainstorm output.}
\label{tab:obj2_metrics}
\small
\begin{tabular}{@{}llp{5.0cm}p{6.2cm}@{}}
\toprule
\textbf{Category} & \textbf{Metric} & \textbf{What It Captures} & \textbf{How It Is Computed} \\
\midrule
\textit{Influence}
  & Mean Idea Influence Depth
  & how many layers of back-and-forth deliberation feed into the typical idea.
  & For each idea, find the longest chain of successive influences leading to it. Average across all ideas in the session. \\
\cmidrule(l){2-4}
  & Mean Cross-Persona Absorption
  & how much of the discussion phase's collaborative output is woven into each idea's lineage.
  & For each idea, count the persona-generated contributions from the discussion phase that appear in its ancestry tree. Average across all ideas, then normalize to 0--100\% of the session-wide maximum. \\
\midrule
\textit{Diversity}
  & Mean Pairwise Idea Distance
  & Semantic spread: how different the session's ideas are from one another.
  & Embed all idea texts, then compute the mean pairwise cosine distance across all idea pairs. \\
\bottomrule
\end{tabular}
\end{table*}

\subsubsection{Cross-Pollination of Disparate Concepts}
The most compelling ideas often arose not from refining a single argument, but from weaving together seemingly unrelated threads. For example, the Busy Working Parent's ``Family Coordination Hub'' idea (Fig.~\ref{fig:output}) drew from multiple conversations at once. Their explanation traces this creative fusion:

\begin{leftbarquote}
{\upshape [Busy Working Parent]:} ``This idea emerged as the discussion moved from tech specs to real household challenges. Foodie's `floury hands, steamy kitchens' constraint became my north star for practical design. The Homebody's locally-processed, privacy-respecting home intelligence concepts---combined with recurring accessibility themes around hands-free use---showed real promise. Travel Enthusiast's cultural context overlays made me realize the same computer vision identifying street signs in other languages could recognize family members at home, transforming domestic mental load into an opportunity for gentle, context-aware assistance that respects both privacy and messy family realities.''
\end{leftbarquote}

This example illustrates how specific, real-world problems from other personas provided powerful grounding metaphors that produced a new idea addressing multiple concerns at once. Fig.~\ref{fig:influence} traces the lineage of one representative idea from this session, illustrating how cross-persona contributions at varying depths converge into a single proposal.

\subsection{Reconciling Idea Generation and Evaluation in System Design}
Our system operationalizes the well-established separation between divergent and convergent thinking through a facilitator-guided approach, in which idea generation (divergent thinking) transitions to idea evaluation (convergent thinking) over time.

\subsubsection{Divergent Thinking (``No Bad Ideas'')}
In the Ideation phase of a session, the facilitator encourages idea submission. Ideas emerge organically from the discussion---any participant may propose one, and no proposal is filtered on the basis of its creative merit. While the system includes an idea-refinement step, its purpose is purely technical: it guards against agentic hallucination and ensures that ideas are coherent and legible, rather than evaluating whether they are good. This preserves the psychological safety that Osborn (1953) identified as essential for creative output---contributors need not fear judgment when there is no quality gate imposed on the substance of their contributions~\cite{b7}.

This early phase generates dynamic, exploratory discourse---high initial variance as ideas are actively challenged and refined rather than passively accepted. Take, for example, the following exchange showing how polite disagreement sharpens an emerging concept:

\begin{leftbarquote}
{\upshape [Weekend Adventurer]:} ``I love the hands-free capture angle, but honestly my first question is about durability and battery life in outdoor settings. When I'm out on a trail for eight hours, I need gear that can handle sweat, rain, temperature swings, and won't die halfway through---plus it needs to work offline since cell service is nonexistent in most places I adventure.''
\end{leftbarquote}

\subsubsection{Convergent Thinking (``Voting as a Ranking Mechanism'')}
As the session progresses, the facilitator gradually shifts emphasis from generation toward evaluation, incentivizing voting over new idea submission. Evaluation is participant-driven: formal votes are deliberate acts of endorsement, and the collective---not the system---determines which ideas gain traction.

The shift from divergent to convergent thinking happens naturally as participants influence and build on each other's ideas. Personas openly adjust their perspectives based on others' input, for example:

\begin{leftbarquote}
{\upshape [Tech Enthusiast]:} ``I'm genuinely reflecting on what the Retired Empty Nester said about technology that enriches experiences versus optimizes them---it's making me realize that my default `build cool tech' instinct sometimes misses the deeper question of whether intelligence actually serves human flourishing.''
\end{leftbarquote}

And dissent, when it arises, tends to deepen rather than derail the conversation:

\begin{leftbarquote}
{\upshape [Accessibility Advocate]:} ``I'm noticing a pattern that concerns me: almost every use case we've discussed assumes visual interaction as the primary mode, which would exclude millions of potential users from day one\ldots{} If we design these glasses with multimodal interaction as the foundation rather than an accessibility afterthought, we create something genuinely innovative that serves everyone better.''
\end{leftbarquote}

Convergence is gained as participants become aligned and vote on the ideas they've generated. Our architecture reconciles what might seem like opposing goals---unconstrained ideation and rigorous evaluation. The facilitator steers the discussion through phases, creating room for free-flowing creativity while guiding the group toward consensus. Crucially, the facilitator manages only the session's arc; qualitative judgments remain with the participants themselves. This separation of responsibilities safeguards the creative freedom that brainstorming demands, while still enabling the group to rally behind its strongest ideas.

\begin{figure}[t]
\centering
\includegraphics[width=\columnwidth]{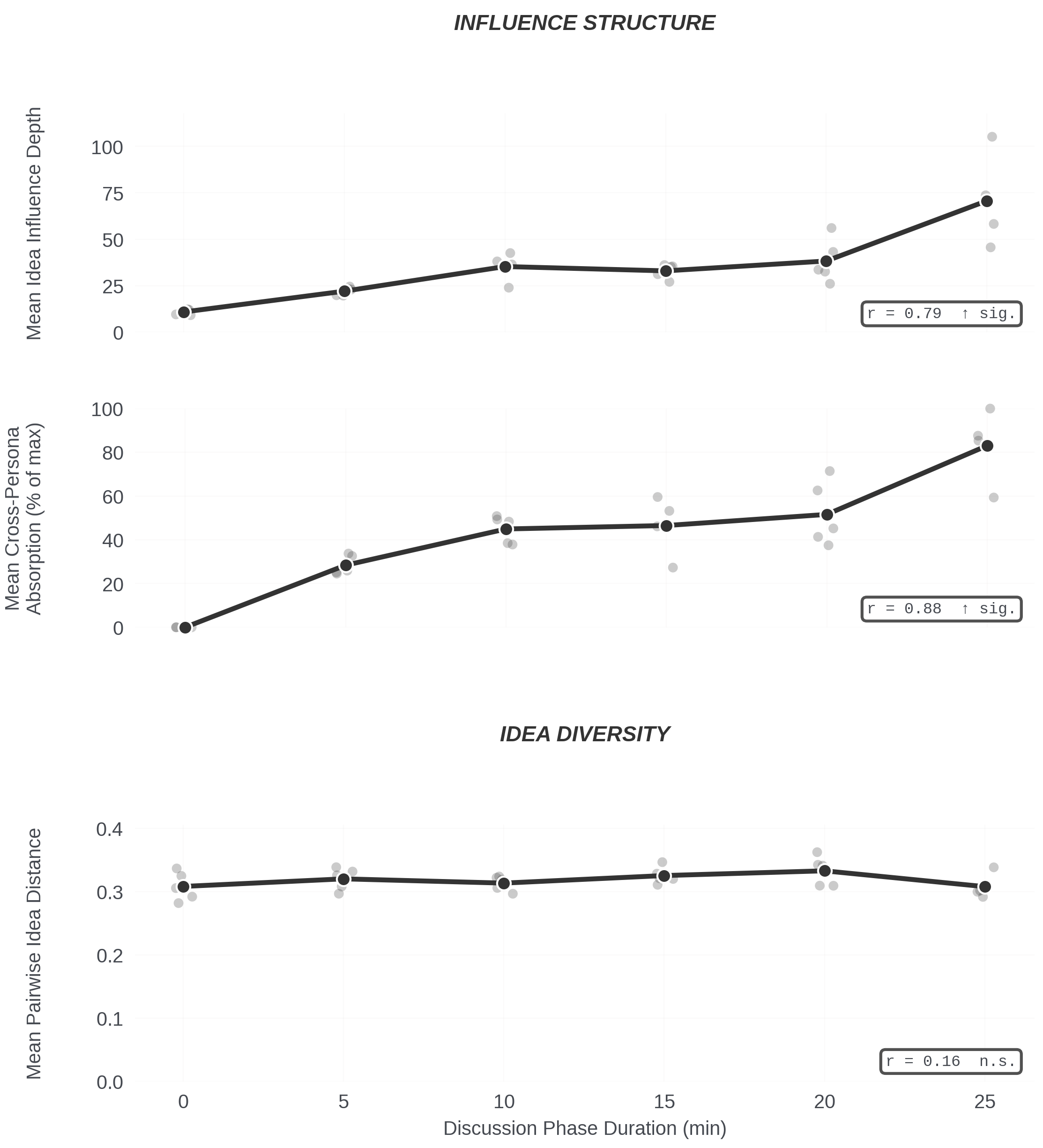}
\caption{Effect of discussion duration on influence structure and idea diversity. Both influence metrics increase significantly with discussion duration, while semantic idea diversity remains relatively flat. Grey dots show individual sessions; black dots show group means. Inset boxes report Pearson correlation $r$ ($\uparrow$~sig.~$= p < 0.05$; n.s.~$=$ not significant).}
\label{fig:Fig_Objective2}
\end{figure}

\section{Objective 2: Does Sustained Discussion Foster Better Ideas?}

To isolate the effect of discussion duration, we conducted 34 independent brainstorming sessions across six durations (0, 5, 10, 15, 20, and 25~minutes), holding all other variables constant (see Section~IV). Two properties of the resulting ideas were measured: \textbf{influence}---the degree to which ideas build on prior contributions from multiple personas---and \textbf{diversity}---the semantic distance between them. Table~\ref{tab:obj2_metrics} defines the metrics that operationalize these constructs, and Figure~\ref{fig:Fig_Objective2} plots each against discussion duration.

\subsubsection{Longer Discussions Produce Deeper Cross-Pollination}

Both influence metrics rose steadily with discussion duration, indicating that longer discussions cause ideas to draw on a deeper lineage of prior contributions. Mean Idea Influence Depth---the length of the longest influence chain behind each idea---increased from 10.8 at 0~minutes to 70.7 at 25~minutes ($r = 0.79$, $p < 0.05$). Mean Cross-Persona Absorption---the proportion of discussion-phase persona contributions present in each idea's ancestry---rose from 0\% to 83\% ($r = 0.88$, $p < 0.05$).

The pattern demonstrates that without discussion, ideas emerge from shallow, largely independent reasoning. With sustained discussion, they crystallize through multi-hop chains of cross-persona deliberation---each comment building on earlier comments from different personas~\cite{bPaulusBrown, b4}.

\subsubsection{Idea Diversity Remains Bounded by Persona Composition}

Yet the idea space itself did not expand significantly with greater discussion. Mean Pairwise Idea Distance remained flat across all durations ($r = 0.16$, n.s.). Ideas produced after 25~minutes of discussion were no more semantically distinct from one another than those generated with no discussion phase at all. Deeper influence chains, it turns out, refine ideas without diversifying them. Conceptual breadth appears to be primarily set by the composition of perspectives in the room---not by how long those perspectives interact~\cite{b9, b16}.

\subsubsection{Implications of Discussion in a Group Setting}

Longer discussions produce significantly deeper reasoning, but not significantly broader ideation. As participants engage, they increasingly build on one another's arguments---refining proposals, challenging assumptions, and articulating tradeoffs. Yet the diversity of the final idea set remains bounded by the personas present. This asymmetry aligns with Steiner's~\cite{b16} distinction between group resources and group process: the \textit{composition} of the group defines the space of possible ideas, while \textit{interaction} determines how rigorously those ideas are examined~\cite{b8, b9}. Discussion duration, then, is best understood as a \textit{depth dial}---a lever for intensifying deliberation, not for expanding the ideational frontier, which the architecture controls structurally through persona selection.

Yet depth and breadth are not entirely independent. Extended deliberation generates a richer shared context~\cite{bCannonBowers}, enabling personas to understand not just \textit{what} was proposed but \textit{why}, what objections arose, and how competing ideas relate to one another. Groups that process shared information more deeply make better decisions, even without generating more options~\cite{bVanKnippenberg}. This accumulated context carries forward into the voting phase, where informed comparison matters more than raw novelty. A group that has deliberated deeply is better equipped to discriminate among its ideas---not because it has generated more of them, but because it has stress-tested the ones it has~\cite{bMercier}. Discussion duration may not widen the pool, but it raises the quality of selection from it.

\section{Applications and Future Work}

The architecture presented here offers a practical tool for early-stage product development. Teams can launch a virtual brainstorming session, crafting personas that mirror a product's specific target audience to pressure-test concepts through the lens of the very customers they aim to serve. The complete lineage of every idea is preserved, enabling teams to trace how audience-specific concerns shaped each proposal and to validate the assumptions embedded within it. Because persona composition is the primary lever for ideational breadth, the same architecture can be retargeted to new domains or market segments simply by recasting the persona pool to reflect a different customer profile.

Several directions remain open for future investigation.

\vspace{2.0ex}
\textit{1) Grounding context in real-world data:} All outputs in this work were generated using off-the-shelf language models without domain-specific fine-tuning. A natural next step is integrating retrieval-augmented generation (RAG) to ground a persona's reasoning in real-world data---such as social sentiment, behavioral signals, or user feedback. This would allow personas to draw on empirical evidence during discussions, bridging the gap between simulated brainstorming and authentic human perspectives.

\vspace{2.0ex}
\textit{2) Optimizing persona generation and composition.}
Our findings show that idea diversity is bounded by the personas in the room rather than by discussion duration, making persona composition the single most consequential design choice. This raises a practical question: what is the optimal number and mix of personas for a given brainstorming objective? Too few may constrain the solution space; too many risk redundancy, weaker differentiation, and diminishing returns.

Underpinning this optimization lies a deeper challenge: while our architecture accepts arbitrary character sketches, encoding a coherent worldview, belief system, set of values, and lived experiences into a concise sketch that reliably steers an agent's reasoning remains an open problem~\cite{b6,b17,b18,bWang}. Advances in persona generation would directly strengthen the architecture presented here, as the quality of brainstorming output is ultimately bounded by the fidelity of the personas that produce it---much like real brainstorming, the best ideas come from having the right people in the room.

\end{document}